\documentclass[sigconf]{aamas} 

\usepackage{balance} 

\setcopyright{none}
\renewcommand\footnotetextcopyrightpermission[1]{}

\title[AAMAS-2022 Formatting Instructions]{Deep Learnable Strategy Templates for Multi-Issue Bilateral Negotiation}

\author{Pallavi Bagga}
\affiliation{
  \institution{Royal Holloway, University of London}
  \city{Egham}
  \country{United Kingdom}}
\email{pallavi.bagga@rhul.ac.uk}

\author{Nicola Paoletti}
\affiliation{
  \institution{Royal Holloway, University of London}
  \city{Egham}
  \country{United Kingdom}}
\email{nicola.paoletti@rhul.ac.uk}

\author{Kostas Stathis}
\affiliation{
  \institution{Royal Holloway, University of London}
  \city{Egham}
  \country{United Kingdom}}
\email{kostas.stathis@rhul.ac.uk}

\begin{abstract}
We study how to exploit the notion of strategy templates to learn strategies for multi-issue bilateral negotiation. Each strategy template consists of a set of interpretable parameterized tactics that are used to decide an optimal action at any time. We use deep reinforcement learning throughout an actor-critic architecture to estimate the tactic parameter values for a threshold utility, when to accept an offer and how to generate a new bid. This contrasts with existing work that only estimates the threshold utility for those tactics. We pre-train the strategy by supervision from the dataset collected using ``teacher strategies'', thereby decreasing the exploration time required for learning during negotiation.  As a result, we build automated agents for multi-issue negotiations that can adapt to different negotiation domains without the need to be pre-programmed. We empirically show that our work outperforms the state-of-the-art in terms of the individual as well as social efficiency.

\end{abstract}


\newcommand{\BibTeX}{\rm B\kern-.05em{\sc i\kern-.025em b}\kern-.08em\TeX}

\newcommand{\param}[1]{{\color{blue}#1}}
\newcommand{\lastobid}{\omega^o_{t}}
\newcommand{\accbid}{\omega^{acc}}
\newcommand{\uthresh}{\bar{u}_t}

\newcommand{\estOppmodel}{\widehat{U}_o}
\newcommand{\ohist}{\Omega^o_t}
\newcommand{\realUsermodel}{U_u}

\makeatletter
\def\@copyrightspace{\relax}
\makeatother
\begin{document}


\pagestyle{fancy}
\fancyhead{}


\maketitle 
\fancyfoot{}
\thispagestyle{empty}

\section{Introduction}
We are concerned with the problem of modelling a self-interested agent negotiating with an opponent over multiple issues while learning to optimally adapt its strategy. 
For instance, an agent trying to buy a laptop, settles the price of a laptop on the behalf of its owner based on a number of other issues such as laptop type, delivery time, payment methods and location delivery~\cite{fatima2006multi}.

For realistic and complex environments, we assume that our agent has no previous knowledge of the opponent's preferences and its negotiating characteristics~\cite{baarslag2016learning}. Also, the utility of offers exchanged during the negotiation decreases over time (in negotiation scenarios with a discount factor), thus, timely decision on rejecting or accepting an offer and making acceptable offers are substantial~\cite{fatima2002multi}. Moreover, in a multi-issue negotiation, there are likely to be a number of different offers at any given utility level. Since they all result in the same utility, our agent is indifferent between these offers. So, there is another challenge to select the best offer which maximizes the utility to the opponent, whilst maintaining our desired utility level (i.e., to aim for the ``win-win'' solution)~\cite{williams2012iamhaggler}.

Existing work consists of four main approaches addressing the above-mentioned challenges. 
(a) Hand-crafted predefined heuristics -- these are proposed in a number of settings with competitive results~\cite{costantini2013heuristic}, and although interpretable (e.g.,~\cite{alrayes2018concurrent, alrayes2014conan}), they are often characterized by ad-hoc parameter/weight settings that are difficult to adapt for different domains. (b) Meta-heuristic (or evolutionary) methods -- work well across domains and improve iteratively using a fitness function (as a guide for quality); however, in these approaches every time an agent decision is made, this needs to be delivered by the meta-heuristic, which is not efficient and does not result in a human-interpretable and reusable negotiation strategy. (c) Machine learning algorithms -- they show the best results with respect to run-time adaptability~\cite{baggaijcai20, razeghi2020deep}, but often their working hypotheses are not interpretable, a fact that may hinder their eventual adoption by users due to lack of transparency in the decision-making that they offer. (d) 
Interpretable strategy templates -- developed in~\cite{bagga2020learnable} to guide the use of a series of tactics whose optimal use can be learned during negotiation. The structure of such templates depends upon a number of learnable choice parameters, determining which acceptance and bidding tactic to employ at any particular time during negotiation. As these tactics represent hypotheses to be tested, defined by the agent developer, they can be explained to a user, and can in turn depend on learnable parameters. 
The outcome of this work is an agent model 
that formulates a strategy template for bid acceptance and generation so that an agent that uses it can make optimal decisions about the choice of tactics while negotiating in different domains~\cite{bagga2020learnable}. 

The benefit of (d) is that it can combine (a), (b) and (c) by using heuristics for the components of the template and meta-heuristics or machine learning for evaluating the choice parameter values of these components. The problem with (d), however, is that the choice parameters of the components for the acceptance and bidding templates are learned once (during training) and used in all the different negotiation settings (during testing) ~\cite{bagga2020learnable}. 
This one-size-fits-all choice of tactics does not accumulate learning experience and may be unsuitable for unknown domains or unknown opponents. In other words, the current mechanism for learning the choice parameter values in~\cite{bagga2020learnable} abstracts away from what is learned in a specific domain once the negotiation has finished, and therefore cannot transfer it to new domains or unseen opponents.

To address the limitation of (d), we propose the idea of using Deep Reinforcement Learning (DRL) to estimate the choice parameter values of components in strategy templates. We name the proposed interpretable strategy templates as ``Deep Learnable Strategy Templates (DLST)''. 
Our contribution is that we study experimentally the ideas behind DLSTs so that agents that employ them to learn parameter values from and across negotiation experiences, hence being capable of transferring the knowledge from one domain to the other, or using the experience against one opponent on the other. This approach leads to ``adaptive'' and generalizable strategy templates. 
We also perform extensive evaluation experiments based on the ANAC tournaments~\cite{jonker2007agent} against agents with learning capabilities (readily available in GENIUS~\cite{williams2014overview}) in a variety of domains with different sizes and competitiveness levels~\cite{williams2014overview}, each with two different profiles. The agents used for comparison span a wide range of strategies and techniques\footnote{E.g., \textit{AgreeableAgent2018}- Frequency-based opponent modelling, \textit{AgentHerb}- Logistic Regression, \textit{SAGA} -Genetic Algorithm (GA), \textit{KakeSoba}- Tabu Search, \textit{Rubick}- Gaussian distribution, \textit{Caduceus2016}- Mixture of GA, algorithm portfolio and experts.}.
Empirically, the DLST-based agent negotiation model outperforms existing strategies in terms of individual as well as social welfare utilities.

The remainder of the paper is organized as follows. In Section~\ref{related}, we discuss the previous work related to learning-based multi-issue negotiation. In Section~\ref{settings}, we give a description of negotiation settings considered in this paper. Then, in Section~\ref{dlst}, the proposed DLST-based negotiation model is introduced followed by various methods and methodologies in Section~\ref{methods}. Subsequently, in Section~\ref{exp}, we experimentally evaluate the performance efficiency of the proposed model. We conclude in Section~\ref{conclusion} where we also outline an open problem worth pursuing in the future, as a result of this work.

\section{Related Work}\label{related}
Existing approaches with reinforcement learning have focused on methods such as Tabular Q-learning for bidding ~\cite{bakker2019rlboa} and finding the optimal concession~\cite{yasumura2009acquisition, yoshikawa2008strategy} or DQN for bid acceptance~\cite{razeghi2020deep}, which are not optimal for continuous action spaces. Such spaces, however, are the main focus in this work in order to estimate the threshold target utility value below which no bid is accepted/proposed from/to the opponent agent. Also, in order to perform this effectively, the agents are required to conclude many prior negotiations with an opponent in order to learn the opponent’s behaviour. Consequently, their approach, and reinforcement learning in general, is not appropriate for one-off negotiation with an unknown opponent. The recently proposed adaptive negotiation model in \cite{baggaijcai20, bagga2021anegma} uses DRL 
for continuous action spaces, but their motivation is significantly different to ours. In our work, the agent attempts to predict the tactic choices for acceptance and bidding strategies at any particular time as well as learn the threshold utility which will be used among one of the tactics to be used in acceptance and bidding strategies, while \cite{baggaijcai20, bagga2021anegma} uses DRL for a complete agent strategy while negotiating with multiple sellers concurrently in e-market like scenarios. Moreover, we focus on building the generalized decoupled and interpretable decision component, i.e., separate acceptance and bidding strategies are learned based on interpretable templates containing different tactics to be employed at different times in different domains.
Another closely related multi-issue DRL-based negotiation work has also been seen in~\cite{bagga2021pareto, bagga2020learnable}. Unlike the use of meta-heuristic optimization to learn the strategy parameter values in~\cite{bagga2020learnable, bagga2021pareto} and use it in all the negotiation settings, we use DRL and the strategy parameter values may differ in different negotiation settings. Also, unlike~\cite{bagga2020learnable, bagga2021pareto}, we abstract away from handling the user preference uncertainties and generating the near-Pareto-optimal bids under preference uncertainties.

\section{Negotiation Settings}\label{settings}
As in~\cite{bagga2020learnable}, we assume that our negotiation environment $E$ consists of two agents $A_u$ and $A_o$ negotiating with each other over some domain $D$. A domain $D$ consists of $n$ different independent issues, $D = (I_1, I_2,  \dots I_n)$, with each issue taking a finite set of $k$ possible discrete or continuous values $I_i = (v^i_1, \ldots v^i_{k})$.  
In our experiments, we consider issues with discrete values. An agent's bid $\omega$ is a mapping from each issue to a chosen value (denoted by $c_i$ for the $i$-th issue), i.e., $\omega = (v^1_{c_1}, \ldots v^n_{c_n})$.  
The set of all possible bids or outcomes is called outcome space $\Omega$ s.t. $\omega \in \Omega$. The outcome space is common knowledge to the negotiating parties and stays fixed during a single negotiation session.
\paragraph{Negotiation protocol}
Before the agents can begin the negotiation and exchange bids, they must agree on a negotiation protocol $P$, which determines the valid moves agents can take at any state of the negotiation ~\cite{fatima2005comparative}. Here, we consider the alternating offers protocol \cite{rubinstein1982perfect}, with possible $Actions = \{\mathit{offer}(\omega), \mathit{accept}, \mathit{reject}\}$. 
\begin{figure*}
    \centering
    \includegraphics[scale=0.4]{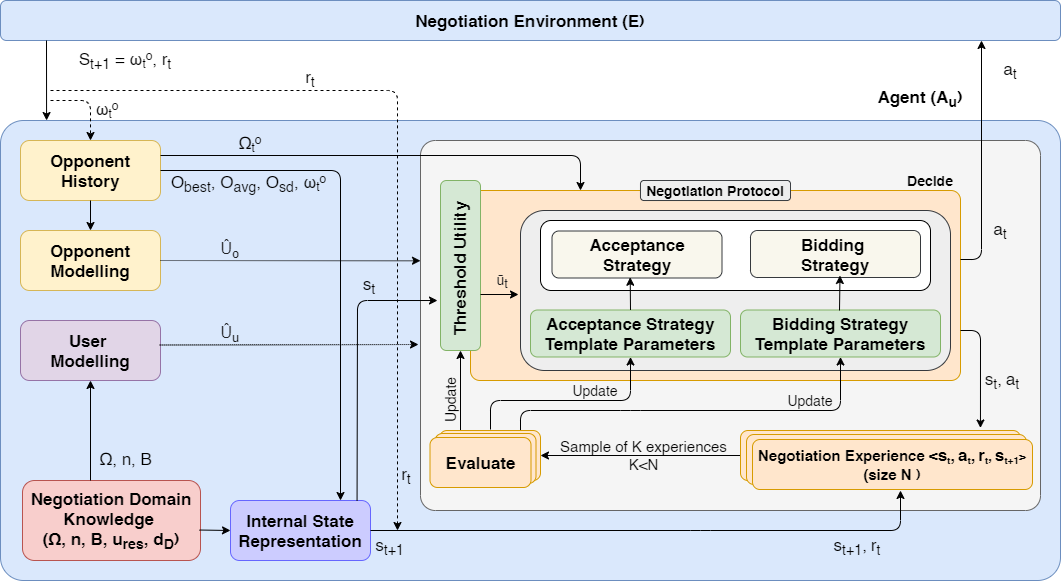}
    \caption{Interaction between the components of DLST-based agent negotiation model}
    \label{anesia-ii}
\end{figure*}
One of the agents (say $A_u$) starts a negotiation by making an offer $x_{A_u \rightarrow A_o}$ to the other agent (say $A_o$). The agent $A_o$ can either accept or reject the offer. If it accepts, the negotiation ends with an agreement, otherwise $A_o$ makes a counter-offer to $A_u$. This process of making offers continues until one of the agents either accepts an offer (i.e., successful negotiation) or the deadline is reached (i.e., failed negotiation).
\paragraph{Time Constraints}
We impose a real–time deadline $t_{end}$ on the negotiation process for both theoretical and practical reasons. The pragmatic reason is that without a deadline, the negotiation might go on forever, especially without any discount factors. Secondly, with unlimited time an agent may simply try a huge amount of proposals to learn the opponent’s preferences~\cite{baarslag2012first}. 
However, taking into account a real–time deadline poses many challenges, such as, agents should be more willing to concede near the deadline, as a break-off yields zero (or the reserved utility, if any) utility for both agents; a real–time deadline also makes it necessary to employ a strategy to decide when to accept an offer; and deciding when to accept involves some prediction whether or not a significantly better opportunity might occur in the future. 

Moreover, we assume that the negotiations are sensitive to \textit{time}, i.e., time impacts the utilities of the negotiating parties. In other words, the value of an agreement decreases over time.
\paragraph{Negotiation session}
Formally, for each negotiation session between two agents $A_u$ and $A_o$, let $a^t_{A_u \rightarrow A_o}  \in Actions$ denote the \textit{offer} action proposed by agent $A_u$ to agent $A_o$ at time $t$. 
A negotiation history $H^t_{A_u \leftrightarrow A_o}$ 
between agents $A_u$ and $A_o$ until time $t$ can be represented as in \eqref{history}:
\begin{equation}
\label{history}
    H^t_{A_u \leftrightarrow A_o}: = (x^{t_1}_{p_1 \rightarrow p_2}, x^{t_2}_{p_3 \rightarrow p_4}, \cdots, x^{t_n}_{p_n \rightarrow p_{n+1}})
\end{equation}
where, $t_n \le t$ and the negotiation actions are ordered over time. Also, $p_j = p_{j+2}$, i.e., the negotiation process strictly follows the alternating-offers protocol. 
Given a negotiation thread
between agents $A_u$ and $A_o$, the action performed by $A_u$ at time $t'$ after receiving an offer $x_{A_o \rightarrow A_u}$ at time $t$ from $A_o$ can be one from the set \textit{Actions} if $t' < t_{end}$, i.e., negotiation deadline is not reached.
Furthermore, we assume bounded rational agents due to the fact that given the limited time, information privacy, and limited computational resources, agents cannot calculate the optimal strategy to be carried out during the negotiation. 

\paragraph{Utility}
We assume that each negotiating agent has its own private preference profile which describes how bids are offered over the other bids. This profile is given in terms of a utility function $U$, defined as a weighted sum of evaluation functions, $e_i(v^i_{c_i})$ as shown in \eqref{utilitySpace}. Each issue is evaluated separately contributing linearly without depending on the value of other issues and hence $U$ is referred to as the Linear Additive Utility space. Here, $w_i$ are the normalized weights indicating the importance of each issue to the user and $e_i(v^i_{c_i})$ is an evaluation function that maps the $v^i_{c_i}$ value of the $i^{th}$ issue to a utility.
\begin{equation}
\label{utilitySpace}
  U(\omega) = U (v^1_{c_1}, \ldots v^n_{c_n}) = \sum^n_{i=1} w_i \cdot e_i(v^i_{c_i}), \text{ where } \sum^n_{i=1} w_i = 1
\end{equation}

Whenever the negotiation terminates without any agreement, each negotiating party gets its corresponding utility based on the private reservation\footnote{The reservation value is the minimum acceptable utility for an agent. It may vary for different parties and different domains. In our settings, it is the same for both parties.} value ($u_{res}$). In case the negotiation terminates with an agreement, each agent receives the discounted utility of the agreed bid, i.e., $U^d(\omega) = U (\omega)d_D^t$. Here, $d_D$ is a discount factor in the interval $[0,1]$ and $t \in [0,1]$ is current normalized time.

\section{DLST-Based Negotiation Model}\label{dlst}
When building a negotiation agent, we normally consider three phases: \textit{pre-negotiation phase} (i.e., estimation of agent owner's preferences, preference elicitation), \textit{negotiation phase} (i.e., offer generation, opponent modelling) and \textit{post-negotiation phase} (i.e., assessing the optimality of offers)~\cite{kiruthika2020lifecycle}. In this paper, we are interested in the second phase, which involves a \textit{Decide} component for choosing an optimal action $a_t$, i.e., $H^{t-1}_{A_o \rightarrow A_u}$.  As in~\cite{bagga2020learnable}, we assume that our agent $A_u$ is situated in an environment $E$ (containing the opponent agent $A_o$) where, at any time $t$, $A_u$ senses the current state $S_t$ of $E$ and represents it as a set of internal attributes, as shown in Figure~\ref{anesia-ii}; however this component is implicit in~\cite{bagga2020learnable}. For the estimation of threshold utility, the set of state attributes include information derived from the sequence of previous bids offered by $A_o$  (e.g., utility of the most recently received bid from the opponent $\omega^o_t$, utility of the best opponent bid so far $O_{best}$, average utility of all the opponent bids $O_{avg}$ and their variability $O_{sd}$) and information stored in $A_u$'s knowledge base (e.g., number of bids $B$ in the given partial order, $d_D$, $u_{res}$, $\Omega$, and $n$), and the current negotiation time $t$. This internal state representation, denoted with $s_t$, is used by the agent (in acceptance and bidding strategies) to decide what action $a_t$ to execute from the set of \textit{Actions} based on the negotiation protocol $P$ at time $t$. Action execution then changes the state of the environment to $S_{t+1}$.
The state $s_t$ for acceptance strategy involves the following attributes in addition to the above-mentioned state attributes: fixed target utility $u$, dynamic and learnable target utility $\bar{u}_t$, $U(\omega)$, $q$ quantile value which changes w.r.t time $t$, $Q_{\widehat{U}(\ohist)}(q)$. On the other hand, the state $s_t$ for bidding strategy involves the following set of attributes: $b_{\mathit{Boulware}}$, $PS$ Pareto-optimal bid, $b_{opp}(\lastobid)$, $\mathcal{U}(\Omega_{\geq \uthresh})$ (as discussed in the subsequent section), in addition to the state attributes used for estimating the dynamic threshold utility value.

The action $a_t$ is derived via two functions, {$f_a$} and {$f_b$}, for the acceptance and bidding strategies, respectively, as in ~\cite{bagga2020learnable}. 
The function $f_a$ takes as inputs $s_t$, a \textit{dynamic threshold utility} $\bar{u}_t$ (defined later in the Methods section), the sequence of past opponent bids $\ohist$, and outputs a discrete action $a_t$ among \textit{accept} or \textit{reject}. When $f_a$ returns \textit{reject}, $f_b$ computes what to bid next, with input $s_t$ and $\bar{u}_t$, see~(\ref{eq:acceptance}--\ref{eq:bidding}). 
This separation of acceptance and bidding strategies is not rare, see for instance~\cite{baarslag2014decoupling}. Also, $f_a$ and $f_b$ consists of a set of tactics as defined in~\cite{bagga2020learnable}.
\begin{align}
    f_a(s_t, \bar{u}_t, \ohist) = & \ a_t, a_t \in \{\mathit{accept, reject}\}\label{eq:acceptance}\\
        f_b(s_t,\bar{u}_t, \ohist) = & \ a_t, a_t \in \{\mathit{offer}(\omega), \omega \in \Omega \}
        \label{eq:bidding}
\end{align}
We assume incomplete opponent preference information, therefore, \textit{Decide} uses the estimated model $\estOppmodel$. In particular, $\estOppmodel$ is estimated at time $t$ using information from $\ohist$, see Methods section for more details. 
 Unlike~\cite{bagga2020learnable}, we employ DRL in Acceptance strategy templates as well as Bidding Strategy templates in our work, in addition to Threshold utility (represented by three green coloured boxes in Figure~\ref{anesia-ii}) in \textit{Decide} component. Each DRL component is actor-critic architecture-based~\cite{sutton2018reinforcement} and has its own \textit{Evaluate} and \textit{Negotiation Experience} components.

\textit{Evaluate} refers to a critic helping our agent learn the dynamic threshold utility $\bar{u}_t$, acceptance strategy template parameters and bidding strategy template parameters, with the new experience collected during the negotiation against each opponent agent. 
More specifically, it is a function of random $K$ ($K<N$) experiences fetched from the agent's memory.
Here, learning is \textit{retrospective}, since it depends on the {reward} $r_t$ obtained from $E$ by performing $a_t$ at $s_t$. 
The reward values for every critic that are used for estimating the threshold utility (i.e., $r_t^{\uthresh}$ ) as well as choice parameter values of acceptance (i.e.,  $r_t^{bid}$) and bidding strategy templates (i.e.,  $r_t^{acc}$) depend on the discounted user utility of the last bid received from the opponent, $\lastobid$, or of the bid accepted by either parties $\accbid$ and defined as \eqref{reward}, \eqref{rewardBidding} and \eqref{rewardAcceptance} respectively.
\begin{equation}\label{reward}
    r_t^{\uthresh} = 
    \begin{cases}
    \realUsermodel(\accbid,t), & \text{on agreement} \\
    \realUsermodel(\lastobid,t), & \text{on received offer} \\
    -1, & \text{otherwise}.
    \end{cases}
\end{equation}
\begin{equation}\label{rewardBidding}
    r_t^{bid} = 
    \begin{cases}
    \realUsermodel(\accbid,t), & \text{on agreement} \\
    -1, & \text{otherwise}.
    \end{cases}
\end{equation}


\begin{equation}\label{rewardAcceptance}
    r_t^{acc} = 
    \begin{cases}
    \realUsermodel(\accbid,t), & \text{on agreement and }U_o(\accbid,t) \leq \realUsermodel(\accbid,t) \\
    \realUsermodel(\lastobid,t), & \text{on rejection and }U_o(\lastobid,t) \geq \realUsermodel(\lastobid,t) \\
    -1, & \text{otherwise}.
    \end{cases}
\end{equation}

$r_t^{\uthresh}$~\eqref{reward} and  $r_t^{bid}$~\eqref{rewardBidding} are straight-forward. In~\eqref{rewardAcceptance}, $U_o(\omega,t)$ is used as the reward value because reward is received from the environment $E$ where the opponent agent resides. In other words, we assume that $E$ has access to $A_o$'s real preferences, i.e., $U_o$, but these preferences are not observable by our agent $A_u$. 
The first case of the $r_t^{acc}$ deals with an agreed bid and returns a positive reward value, if the bid gives higher utility to our agent than the opponent. The second case deals with a rejected bid and returns a positive reward value, if the bid gives lower utility to our agent than the opponent. In all other cases, it returns a negative value.
Also, in~\eqref{reward}, \eqref{rewardBidding} and \eqref{rewardAcceptance}, $\realUsermodel(\omega,t)$ is the discounted reward of $\omega$ defined as~\eqref{utility}. 
\begin{equation}\label{utility}
   \realUsermodel(\omega,t) =\realUsermodel(\omega)\cdot{d^t}, d \in [0,1]
\end{equation}
In~\eqref{utility}, $d$ is a temporal discount factor to encourage the agent to negotiate without delay. 
We should not confuse $d$, which is typically unknown to the agent, with the discount factor used to compute the utility of an agreed bid ($d_D$).

\textit{Negotiation Experience} stores historical information about $N$ previous interactions of an agent with other agents. Experience elements are of the form $\langle s_t, a_t, r_t, s_{t+1} \rangle$, where $s_t$ is the internal state representation of the negotiation environment $E$, $a_t$ is the performed action, $r_t$ is a scalar \textit{reward} received from the environment and $s_{t+1}$ is the new agent state after executing $a_t$. 

\paragraph{Strategy templates}
The strategy templates of~\cite{bagga2020learnable} are a general form of parametric strategies for acceptance and bidding. These strategies apply different tactics at different phases of the negotiation. The total number of phases $n$ and the number of tactics $n_i$ to choose from at each phase $i=1,\ldots,n$ are the only parameters fixed in advance. For each phase $i$, the duration $\delta_i$ (i.e., $t_{i+1} = t_{i} + \delta_i$) and the particular choice of tactic are learnable parameters. The latter is encoded with choice parameters $c_{i,j}$, where $i=1,\ldots,n$ and  $j=1,\ldots,n_i$, such that if $c_{i,j}$ is true then the $(i,j)$-th tactic is selected for phase $i$. Tactics can be parametric in turn, and depend on learnable parameters $\mathbf{p}_{i,j}$. 

We consider the same set of admissible tactics as~\cite{bagga2020learnable}. The key difference is that our approach allows to evolve the entire strategy (within the space of strategies entailed by the template) at every negotiation, which makes more adaptable and generalizable. The tactics used for acceptance strategies are:
\begin{itemize}
\item $\realUsermodel(\omega_t)$, the estimated utility of the bid $\omega_t$ that our agent would propose at time $t$. 
\item 
$Q_{\realUsermodel(\ohist)}(\param{a}\cdot t + \param{b})$, where $\realUsermodel(\ohist)$ is the distribution of (estimated) utility values of the bids in $\ohist$, $Q_{\realUsermodel(B_o(t))}(p)$ is the quantile function of such distribution, and $\param{a}$ and $\param{b}$ are learnable parameters. In other words, we consider the $p$-th best utility received from the agent, where $p$ is a learnable (linear) function of the negotiation time $t$. In this way, this tactic automatically and dynamically decides how much the agent should concede at time $t$. Here, $\mathbf{p}_{i,j} = \{a, b\}$ .
\item $\uthresh$, the dynamic DRL-based utility threshold.
\item $u$, a fixed utility threshold.
\end{itemize}
The bidding tactics are:
\begin{itemize}
\item $b_{\mathit{Boulware}}$, a bid generated by a time-dependent Boulware strategy~\cite{fatima2001optimal}.
\item $PS(\param{a}\cdot t + \param{b})$ extracts a bid from the set of Pareto-optimal bids $PS$, derived using the \textit{NSGA-II algorithm}\footnote{Meta-heuristics (instead of brute-force) for Pareto-optimal solutions have the potential to deal efficiently with continuous issues.}~\cite{deb2002fast} under $\realUsermodel$ and $\estOppmodel$.
In particular, it selects the bid that assigns a weight of $\param{a}\cdot t + \param{b}$ to our agent utility (and $1-(\param{a}\cdot t + \param{b})$ to the opponent's), where $\param{a}$ and $\param{b}$ are learnable parameters telling how this weight scales with the negotiation time $t$. The \textit{TOPSIS algorithm}~\cite{hwang1981methods}  is used to derive such a bid, given the weighting $\param{a}\cdot t + \param{b}$ as input. Here, $\param{\mathbf{p}_{i,j} = \{a, b\}}$ .
\item $b_{opp}(\lastobid)$, a tactic to generate a bid by manipulating the last bid received from the opponent $\lastobid$. This is modified in a greedy fashion by randomly changing the value of the least relevant issue (w.r.t.\ ${U}$) of $\lastobid$. \item $\omega \sim \mathcal{U}(\Omega_{\geq \uthresh})$, a random bid above our DRL-based utility threshold $\uthresh$\footnote{$\mathcal{U}(S)$ is the uniform distribution over $S$, and $\Omega_{\geq \uthresh}$ is the subset of $\Omega$ whose bids have estimated utility above $\uthresh$ w.r.t.\ ${U}$.}.
\end{itemize}

\noindent Below, we give an example of a concrete acceptance strategy learned with our model. We use, as we will discuss in Section~\ref{exp}, a specific domain (\textit{Party}) and we show how the strategy adapts in other negotiation domains (\textit{Grocery} and \textit{Outfit}) against the opponent strategy~\cite{bagga2020learnable}.\\

\begin{small}
\noindent(a) Party Domain
\begin{align*}
t \in [0.000, 0.0361) \rightarrow & \
 U_u(\omega_t^o) \geq \max\left(Q_{U_{\Omega^o_t}} (-0.20 \cdot t + 0.22), \bar{u_t}\right)\\
t \in [0.0361, 1.000] \rightarrow &  \
U_u(\omega_t^o) \geq \max\left( u, Q_{U_{\Omega^o_t}} (-0.10 \cdot t + 0.64)\right)
\end{align*} 
(b) Grocery Domain
\begin{align*}
t \in [0.000, 0.2164) \rightarrow & \
U_u(\omega_t^o) \geq \max\left( U_u(\omega_t), Q_{U_{\Omega^o_t}} (-0.55 \cdot t + 0.05),\bar{u_t} \right) \\
 t \in [0.2164, 0.3379) \rightarrow &  \
U_u(\omega_t^o) \geq \max\left( U_u(\omega_t), Q_{U_{\Omega^o_t}} (-0.60 \cdot t + 1.40)\right)\\
t \in [0.3379, 1.000] \rightarrow &  \
U_u(\omega_t^o) \geq \max\left( Q_{U_{\Omega^o_t}} (-0.22 \cdot t + 0.29), \bar{u_t}\right)
\end{align*} 
(c) Outfit Domain
\begin{align*}
t \in [0.000, 0.1545) \rightarrow & \
U_u(\omega_t^o) \geq Q_{U_{\Omega^o_t}} (-0.50 \cdot t + 0.70)\\
 t \in [0.1545, 0.3496) \rightarrow &  \
U_u(\omega_t^o) \geq \max\left( \bar{u_t} , Q_{U_{\Omega^o_t}} (-0.50 \cdot t + 0.90)\right)\\
t \in [0.3496, 1.000] \rightarrow &  \
U_u(\omega_t^o) \geq U_u(\omega_t)
\end{align*} 
\end{small}
We can observe that the duration learned in the left-hand side of the tactics is different for different domains, e.g., initially in the first domain ($Party$) the first rule triggers when $t \in [0.0, 0.0361)$, while in the second ($Grocery$) and third ($Outfit$) domains, the first rule triggers at  $t \in [0.0, 0.2164)$ and $t \in [0.0, 0.1545)$ respectively. Similarly, the parameters on the right-hand side of the tactics rules, e.g., for the first domain ($Party$) during the very early phase of the negotiation, the strategy uses a quantile tactic as well as dynamic threshold utility. However, in the second domain ($Grocery$), the strategy now employs future bid utility along with the quantile bid and the dynamic threshold utility tactics, whereas, in the third domain ($Outfit$), it only employs the quantile bid tactic.



\section{Methods}\label{methods}
In our approach, we first use supervised learning (SL) to pre-train the our agent using supervision examples collected from existing ``teacher'' negotiation strategies as inspired by~\cite{bagga2020learnable, bagga2021anegma}. Such pre-trained strategy is then evolved via RL using experience and rewards collected while interacting with other agents in the negotiation environment.
This combination of SL and RL approaches enhances the process of learning an optimal strategy. This is because applying RL alone from scratch would require a large amount of experience before reaching a reasonable strategy, which might hinder the online performance of our agent. On the other hand, starting from a pre-trained policy ensures quicker convergence (as demonstrated empirically in~\cite{bagga2020learnable, bagga2021anegma}). 

\subsection{Data set collection}
In order to collect the data set for pre-training our agent via SL, we have used the \textit{GENIUS} simulation environment~\cite{lin2014genius}. In particular, in our experiments we generate supervision data using the existing DRL-based state-of-the-art agent negotiation model~\cite{bagga2020learnable} 
by negotiating it against the winning strategies of ANAC-2019 competition, i.e., AgentGG, KakeSoba and SAGA (readily available in GENIUS and requiring minimal changes to work for our negotiation settings) assuming no user preference uncertainty in three different domains (Laptop, Holiday, and Party).

\subsection{Strategy Representation}
We represent both $f_a$ \eqref{eq:acceptance} and $f_b$ \eqref{eq:bidding} using artificial neural networks (ANNs)~\cite{goodfellow2016deep}, as these are powerful function approximators and benefit from extremely effective learning algorithms, unlike~\cite{bagga2020learnable}, which used the meta-heuristic optimization algorithm. We also use the same to predict the target threshold utility $\uthresh$ as in~\cite{bagga2020learnable}.
\subsubsection{ANN}
In particular, we use feed-forward neural networks, i.e., functions organized into several layers, where each layer comprises a number of neurons that process information from the previous layer. More details can be found in~\cite{goodfellow2016deep}. Also, we keep the ANN configuration same as in~\cite{bagga2020learnable}. 
\subsubsection{DRL}
During our experiments, the agent negotiates with fixed-but-unknown opponent strategies in a negotiation environment, which can be learnt by our agent after some simulation runs. Hence, we consider our negotiation environment as \textit{fully-observable}. Following this, for our \textit{dynamic} and \textit{episodic} environment, we use a \textit{model-free}, \textit{off-policy} RL approach which generates a \textit{deterministic policy} based on the \textit{policy gradient} method to support continuous control. More specifically, as in~\cite{bagga2020learnable}, we use Deep Deterministic Policy Gradient (DDPG) algorithm, which is an actor-critic RL approach and generates a deterministic action selection policy for the negotiating agent~\cite{lillicrap2017continuous}. We consider a \textit{model-free} RL approach because our problem is how to make an agent decide what action to take next in a negotiation dialogue rather than predicting the new state of the environment. In other words, we are not learning a model of the environment, as the strategies of the opponents are not observable properties of the environment's state. Thus, our agent's emphasis is more on learning what action to take next and not the state transition function of the environment. We consider the \textit{off-policy} approach (i.e., an agent attempts to evaluate or improve the policy which is different from the one which was used to take an action) for independent exploration of continuous action spaces~\cite{lillicrap2017continuous}.
When being in a state $s_t$, DDPG uses a so-called \textit{actor} network $\mu$ to select an action $act_t$, and a so-called \textit{critic} network $Q$ to predict the value $Q_t$ at state $s_t$ of the action selected by the actor:
\begin{align}
    act_t = & \ \mu(s_t \mid \theta^\mu) \label{actor}\\
    Q_t(s_t,act_t \mid \theta^Q) = & \ Q(s_t,\mu(s_t \mid \theta^\mu )\mid\theta^Q)\label{critic}
\end{align}
In \eqref{actor} and \eqref{critic}, $\theta^\mu$ and $\theta^Q$ are, respectively, the learnable parameters of the actor and critic neural networks. The parameters of the actor network are updated by the Deterministic Policy Gradient method~\cite{silver2014deterministic}. The objective of the actor policy function is to maximize the expected return $J$ calculated by the critic function using~\eqref{return}. See~\cite{lillicrap2015continuous} for further details on DDPG.
\begin{equation}\label{return}
    J = \mathbb{E}[Q(s, act|\theta^Q)|_{s=s_t, act=\mu(s_t)}]
\end{equation}

In our experiments, for predicting the dynamic threshold utility, the actor function is a single-output regression ANN; on the other hand, for acceptance and bidding strategies, it is a multiple-output regression ANN. In particular, when predicting $\uthresh$, $act_t$ corresponds to $\uthresh$; whereas, for acceptance and bidding strategy templates, $act_t$ consists of a vector of multiple outputs $\left(\param{\delta_i}, (\param{c_{i,j}}, \param{\mathbf{p}_{i,j}})_{j=1,\ldots,n_i}\right)_{i=1,\ldots,n}$ including the duration of each negotiation phase $\delta_i$, Boolean choice parameters $c_{i,j}$ and a set of learnable parameters $\mathbf{p}_{i,j}$ for each tactic $j$ that can be used in a negotiation phase $i$. 

\subsection{Opponent modelling}
We consider a negotiation environment with uncertainty about the opponent's preferences. 
To derive an estimate of the opponent model $\estOppmodel$ during negotiation, 
we use the distribution-based frequency model proposed in~\cite{tunali2017rethinking}, as also done in~\cite{bagga2020learnable}.
In this model, the empirical frequency of the issue values in $\ohist$ provides an educated guess on the opponent's most preferred issue values. 
The issue weights are estimated by analysing the disjoint windows of $\ohist$, giving an idea of the shift of opponent's preferences from its previous negotiation strategy over time.


\section{Experimental Results and Discussions}\label{exp}
All the experiments are performed using the  GENIUS tool~\cite{lin2014genius}, which are designed to prove the following two hypotheses:
\begin{itemize}
    \item\textbf{Hypothesis A:} DLST-based negotiation approach outperforms the ``teacher'' strategies in known negotiation settings in terms of individual and social efficiency.

    \item\textbf{Hypothesis B:} DLST-based negotiation approach outperforms not-seen-before strategies and adapts to different negotiation settings in terms of individual and social efficiency.
    
\end{itemize}

\subsection{Performance metrics:}
We measure the performance of each agent in terms of six widely-adopted metrics inspired by the ANAC competition:
\begin{itemize}
    \item $U_{\it ind}^{total}$: The utility gained by an agent averaged over all the negotiations ($\uparrow$);
    \item $U_{\it ind}^s$: The utility gained by an agent averaged over all the \textit{successful} negotiations ($\uparrow$);    
    \item $U_{\it soc}$: The utility gained by both negotiating agents averaged over all successful negotiations ($\uparrow$);
    \item $P_{\it avg}$: Average minimal distance of agreements from the Pareto Frontier ($\downarrow$).
    \item $S_{\%}$: Proportion of successful negotiations ($\uparrow$).
\end{itemize}
The first and second measures represent \textit{individual efficiency} of an outcome, whereas the third and fourth correspond to the \textit{social efficiency} of agreements. 

\subsection{Experimental settings}  Our proposed DLST-based agent negotiation model is evaluated against state-of-the-art strategies that participated in ANAC'17 and ANAC'18, which are designed by different research groups independently. Each agent has no information about another agent's strategies beforehand. Details of all these strategies are available in~\cite{aydougan2018anac, jonker2017automated}.
We evaluate our approach on total of $11$ negotiation domains which are different from each other in terms of size and opposition~\cite{baarslag2013evaluating} to ensure good negotiation characteristics and to reduce any biases. The domain size refers to the number of issues, whereas opposition\footnote{The value of opposition reflects the competitiveness between parties in the domain. Strong opposition means a gain of one party is at the loss of the other, whereas, weak opposition means that both parties either lose or gain simultaneously~\cite{baarslag2013evaluating}.} refers to
the minimum distance from all possible outcomes to the point representing complete satisfaction of both negotiation parties (1,1). 
For the experiments of Hypothesis B, we choose readily-available 3 small-sized, 2 medium-sized, and 3 large-sized domains. Out of these domains, 2 are with high, 3 with medium and 3 with low opposition (see~\cite{williams2014overview} for more details). 

For each configuration, each agent plays both roles in the negotiation (e.g., buyer and seller in Laptop domain) to compensate for any utility differences in the preference profiles. We call \textit{user profile} the agent's role along with the user's preferences. Also, we set the $u_{res}$ and $d_D$ to their respective default values, whereas the deadline is set to 180s, normalized in $[0,1]$ (known to both negotiating parties in advance). 
For NSGA-II during the Pareto-bid generation phase, we choose the population size of $2\%\times|\Omega|$, $2$ generations and mutation count of $0.1$.
With these hyperparameters, on our machine\footnote{CPU: 8 cores, 2.10GHz; RAM: 32 GB} the run-time of NSGA-II never exceeded the given timeout of 10s for deciding an action at each turn, while being able to retrieve empirically good solutions. 

\subsection{Empirical Evaluation}
We evaluate and discuss the two hypotheses introduced at the beginning of the section.
\subsubsection{Hypothesis A: DLST-based agent outperforms ``teacher'' strategies}
We performed a total of $1200$ negotiation sessions\footnote{$n \times (n-1)/2 \times x \times y \times z = 1200$ where $n = 5$, number of agents in a tournament; $x=2$, because agents play both sides; $y=3$, number of domains; $z=20$, because each tournament is repeated 20 times.}
to evaluate the performance of \textit{DLST-based agent} against the four ``teacher'' strategies (ANESIA~\cite{bagga2020learnable}, AgentGG, KakeSoba and SAGA) in three domains (Laptop, Holiday, and Party). These strategies were used to collect the dataset in the same domains for supervised training before the DRL process begins. Table~\ref{table:teacher} demonstrates the average results over all the domains and profiles for each agent. Clearly,  \textit{DLST-based agent} outperforms the ``teacher'' strategies in terms of  individual efficiency, as well as social efficiency.
\begin{table*}
    \centering
    \begin{tabular}{p{0.13\textwidth}p{0.15\textwidth}
    p{0.15\textwidth}p{0.15\textwidth}
    p{0.15\textwidth}p{0.06\textwidth}}
    \hline
      Agent &
      $P_{\it avg} (\downarrow)$  &  
      $U_{\it soc} (\uparrow)$  &  
      $U_{\it ind}^{total} (\uparrow)$  &  
      $U_{\it ind}^s (\uparrow)$  &
      $S_{\%} (\uparrow)$ \\  
      \hline
      \multicolumn{6}{c}{Laptop Domain} \\
        \hline
    DLST-agent & \textbf{0.0 $\pm$ 0.0} & \textbf{1.71 $\pm$ 0.03} & \textbf{0.91 $\pm$ 0.02} &\textbf{0.91 $\pm$ 0.02} & \textbf{1.00} \\
    ANESIA & \textbf{0.0 $\pm$ 0.0} & 1.66 $\pm$ 0.20 & 0.86 $\pm$ 0.03 & 0.86 $\pm$ 0.03 & \textbf{1.00 }\\
    KakeSoba & 0.03 $\pm$ 0.12 & 1.48 $\pm$ 0.53 & 0.77 $\pm$ 0.20 & 0.82 $\pm$ 0.06 & 0.94 \\
    SAGA & 0.01 $\pm$ 0.06 & 1.45 $\pm$ 0.48 & 0.89 $\pm$ 0.13 & 0.89 $\pm$ 0.10 & 0.99 \\
    AgentGG* & 0.22 $\pm$ 0.35 & 1.14 $\pm$ 0.65 & 0.71 $\pm$ 0.38 & \textbf{0.91 $\pm$ 0.09} & 0.78 \\

         \hline
         \multicolumn{6}{c}{Holiday Domain} \\
           \hline
    DLST-agent & \textbf{0.05 $\pm$ 0.11} & \textbf{1.74 $\pm$ 0.14 }&\textbf{0.96 $\pm$ 0.14} & \textbf{0.96 $\pm$ 0.14} & \textbf{1.00 }\\
    ANESIA & 0.06 $\pm$ 0.1 & \textbf{1.74 $\pm$ 0.14} & 0.85 $\pm$ 0.15 & 0.85 $\pm$ 0.15 & \textbf{1.00} \\
    KakeSoba & 0.21 $\pm$ 0.35 & 1.53 $\pm$ 0.5 & 0.84 $\pm$ 0.27 & 0.92 $\pm$ 0.07 & 0.91 \\
    SAGA & 0.19 $\pm$ 0.36 & 1.55 $\pm$ 0.5 & 0.70 $\pm$ 0.25 & 0.77 $\pm$ 0.12 & 0.91 \\
    AgentGG* & 0.46 $\pm$ 0.58 & 1.16 $\pm$ 0.82 & 0.74 $\pm$ 0.45 & \textbf{0.96 $\pm$ 0.03} & 0.67 \\

         \hline
         \multicolumn{6}{c}{Party Domain} \\
           \hline
    DLST-agent & \textbf{0.15 $\pm$ 0.38} & \textbf{1.53 $\pm$ 0.6} &\textbf{0.74 $\pm$ 0.31} & \textbf{0.77 $\pm$ 0.14} & \textbf{0.87 }\\
    ANESIA & 0.37 $\pm$ 0.32 & 1.06 $\pm$ 0.5 & 0.52 $\pm$ 0.27 & 0.62 $\pm$ 0.14 & 0.83 \\
    KakeSoba & 0.33 $\pm$ 0.32 & 1.11 $\pm$ 0.51 & 0.64 $\pm$ 0.3 & 0.75 $\pm$ 0.12 & 0.84 \\
    SAGA & \textbf{0.15 $\pm$ 0.16} & 1.36 $\pm$ 0.26 & 0.61 $\pm$ 0.19 & 0.63 $\pm$ 0.16 & \textbf{0.87} \\
    AgentGG* & 0.38 $\pm$ 0.42 & 0.92 $\pm$ 0.6 & 0.62 $\pm$ 0.4 & \textbf{0.77 $\pm$ 0.12} & 0.71 \\

         \hline
    \end{tabular}
    \caption{Performance Comparison of \textit{DLST-agent} with ``teacher'' strategies for all the three domains (Laptop, Holiday, and Party - All readily available in GENIUS). Best Results are in \textbf{bold}. Note\textbf{ *} means user preference uncertainty is considered.}
    \label{table:teacher}
    \vspace{-4mm}
\end{table*}
\subsubsection{Hypothesis B: Adaptive behaviour of DLST-based agents}
\begin{figure}
    \centering
    \includegraphics[scale=0.6]{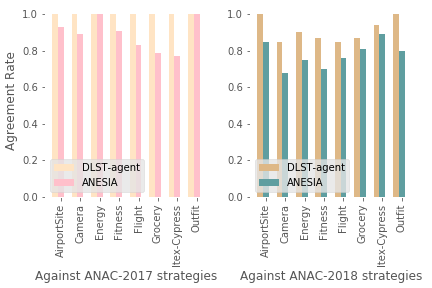}
    \caption{Comparison of DLST-agent VS ANESIA in terms of Agreement rate $S_{\%} (\uparrow)$}
    \label{Fig1}
\end{figure}

\begin{figure}
    \centering
    \includegraphics[scale=0.6]{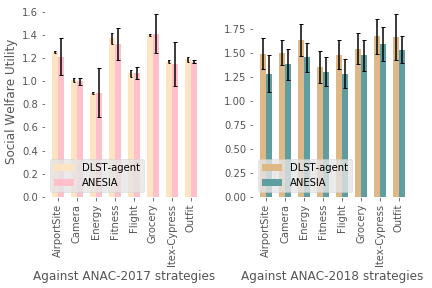}
    \caption{Comparison of DLST-agent VS ANESIA in terms of Social welfare utility $U_{\it soc} (\uparrow)$}
    \label{Fig2}
\end{figure}

\begin{figure}
    \centering
    \includegraphics[scale=0.6]{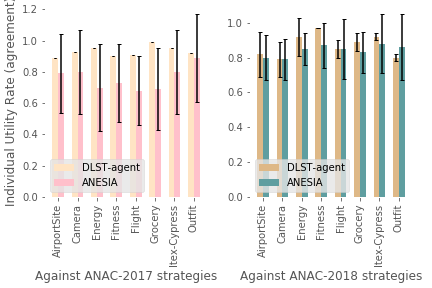}
    \caption{Comparison of DLST-agent VS ANESIA in terms of individual utility rate over successful negotiations  $U_{\it ind}^s (\uparrow)$}
    \label{Fig4}
\end{figure}

\begin{figure}
    \centering
    \includegraphics[scale=0.6]{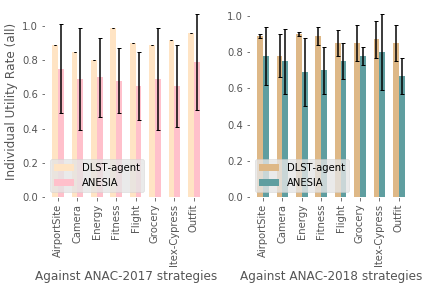}
    \caption{Comparison of DLST-agent VS ANESIA in terms of individual utility rate over all negotiations  $U_{\it ind}^{total}(\uparrow)$ }
    \label{Fig5}
\end{figure}

We further evaluated the performance of \textit{DLST-based agent} against the opponent agents from ANAC'17 and ANAC'18 unseen during training and having capability of learning from previous negotiations. 
For this, we performed two experiments against ANAC'17 and ANAC'18 agents, each with a total of $29120$ negotiation sessions\footnote{$n \times (n-1)/2 \times x \times y \times z = 29120$ where $n = 14$; $x=2$; $y=8$; $z=20$.}.
Results in Table \ref{table:hypB} are averaged over all domains, and  demonstrate that \textit{DLST-based} agent learns to make the optimal choice of tactics to be used at run time and outperforms the other $8$ strategies
in terms of $U_{\it ind}^{s}$ and $U_{\it soc}$. We also observed that our agent outperforms the current state-of-the-art (ANESIA) in a tournament with ANAC'17 and ANAC'18 strategies in all the domains used for the purpose of evaluation as shown in Figures~\ref{Fig1} --~\ref{Fig5}. This indicates that the DLST approach of dynamically adapting the parameters of acceptance and bidding strategies leads consistently improve the ANESIA approach of keeping these parameters fixed once the agent is deployed.
\begin{table*}
    \centering
    \begin{tabular}{p{0.13\textwidth}p{0.15\textwidth}
    p{0.15\textwidth}p{0.15\textwidth}
    p{0.15\textwidth}p{0.06\textwidth}}
    \hline
      Agent &
      $P_{\it avg} (\downarrow)$  &  
      $U_{\it soc} (\uparrow)$  &  
      $U_{\it ind}^{total} (\uparrow)$  &  
      $U_{\it ind}^s (\uparrow)$  &
      $S_{\%} (\uparrow)$ \\  
    
      \hline
 \multicolumn{6}{c}{Comparison of DLST and ANESIA with ANAC 2017 Agent Strategies} \\
   \hline
    DLST-agent &\textbf{0.0 $\pm$ 0.0} &\textbf{1.17 $\pm$ 0.12} & \textbf{0.90 $\pm$ 0.0 }& \textbf{0.93 $\pm$ 0.0} & \textbf{1.0} \\
    ANESIA & \textbf{0.0 $\pm$ 0.0} & 1.16 $\pm$ 0.12 & 0.70 $\pm$ 0.25 & 0.76 $\pm$ 0.26 & 0.89 \\
    PonpokoAgent & 0.70 $\pm$ 0.49 & 0.44 $\pm$ 0.70 & 0.62 $\pm$ 0.19 & \textbf{0.93 $\pm$ 0.04} & 0.89 \\
    ShahAgent & 0.54 $\pm$ 0.54 & 0.79 $\pm$ 0.79 & 0.57 $\pm$ 0.07 & 0.64 $\pm$ 0.04 & 0.75 \\
    Mamenchis & 0.50 $\pm$ 0.05 & 0.80 $\pm$ 0.80 & 0.66 $\pm$ 0.16 & 0.82 $\pm$ 0.18 & 0.89 \\
    AgentKN &\textbf{0.0 $\pm$ 0.0} & \textbf{1.17 $\pm$ 0.0} & 0.65 $\pm$ 0.05 & 0.65 $\pm$ 0.05 & \textbf{1.0} \\
    Rubick & 1.08 $\pm$ 0.0 & 1.00 $\pm$ 0.0 & 0.50 $\pm$ 0.09 & 0.64 $\pm$ 0.04 & 0.76 \\
    ParsCat2 & 0.54 $\pm$ 0.54 & 0.80 $\pm$ 0.08 & 0.66 $\pm$ 0.16 & 0.82 $\pm$ 0.04 & 0.57 \\
    SimpleAgent & 1.08 $\pm$ 0.0 & 0.90 $\pm$ 0.0 & 0.57 $\pm$ 0.14 & 0.57 $\pm$ 0.14 & \textbf{1.0} \\
    AgentF & 1.18 $\pm$ 0.0 & 1.07 $\pm$ 0.06 & 0.51 $\pm$ 0.0 & 0.81 $\pm$ 0.0 & 0.89 \\
    TucAgent & 0.08 $\pm$ 0.29 & 0.90 $\pm$ 0.03 & 0.65 $\pm$ 0.38 & 0.52 $\pm$ 0.16 & 0.69 \\
    MadAgent & 0.67 $\pm$ 0.05 & 1.09 $\pm$ 0.17 & 0.57 $\pm$ 0.0 & 0.57 $\pm$ 0.0 & \textbf{1.0} \\
    GeneKing & 1.08 $\pm$ 0.0 & 0.99 $\pm$ 0.14 & 0.75 $\pm$ 0.0 & 0.67 $\pm$ 0.24 & 0.63 \\
    Farma17 & 0.77 $\pm$ 0.49 & 0.44 $\pm$ 0.70 & 0.65 $\pm$ 0.19 & \textbf{0.93 $\pm$ 0.04} & 0.79 \\

  \hline
         \hline
    \multicolumn{6}{c}{Comparison of DLST and ANESIA with ANAC 2018 Agent Strategies} \\
    \hline
   DLST-agent & \textbf{0.00 $\pm$ 0.08} &\textbf{1.54 $\pm$ 0.17} & \textbf{0.86 $\pm$ 0.07} & \textbf{0.87 $\pm$ 0.06} & \textbf{0.91 }\\
    ANESIA & \textbf{0.00 $\pm$ 0.09} & 1.41 $\pm$ 0.16 & 0.74 $\pm$ 0.14 & 0.84 $\pm$ 0.14 & 0.78 \\
    AgentHerb & 0.02 $\pm$ 0.05 & 0.79 $\pm$ 0.11 & 0.78 $\pm$ 0.02 & 0.78 $\pm$ 0.11 & 0.61 \\
    AgreeableAgent & 0.05 $\pm$ 0.11 & 1.12 $\pm$ 0.23 & 0.53 $\pm$ 0.10 & 0.56 $\pm$ 0.05 & 0.54 \\
    Sontag & 0.03 $\pm$ 0.07 & 0.73 $\pm$ 0.18 & 0.78 $\pm$ 0.08 & 0.79 $\pm$ 0.07 & 0.59 \\
    Agent33 & 0.04 $\pm$ 0.07 & 0.74 $\pm$ 0.18 & 0.68 $\pm$ 0.09 & 0.78 $\pm$ 0.09 & 0.79 \\
    AngentNP1 & 0.04 $\pm$ 0.06 & 0.73 $\pm$ 0.16 & 0.65 $\pm$ 0.10 & 0.65 $\pm$ 0.1 & 0.69 \\
    FullAgent & 0.02 $\pm$ 0.04 & 0.67 $\pm$ 0.12 & 0.69 $\pm$ 0.05 & 0.77 $\pm$ 0.12 & 0.61 \\
    ATeamAgent & 0.09 $\pm$ 0.06 & 0.58 $\pm$ 0.13 & 0.75 $\pm$ 0.10 & 0.75 $\pm$ 0.08 & 0.75 \\
    ConDAgent & 0.06 $\pm$ 0.09 & 1.16 $\pm$ 0.20 & 0.68 $\pm$ 0.11 & 0.65 $\pm$ 0.11 & 0.56 \\
    GroupY & 0.03 $\pm$ 0.06 & 0.66 $\pm$ 0.15 & 0.53 $\pm$ 0.07 & 0.54 $\pm$ 0.06 & 0.58 \\
    Yeela & 0.04 $\pm$ 0.06 & 0.68 $\pm$ 0.14 & 0.73 $\pm$ 0.08 & 0.73 $\pm$ 0.07 & 0.66 \\
    Libra & 0.10 $\pm$ 0.09 & 0.54 $\pm$ 0.19 & 0.71 $\pm$ 0.08 & 0.56 $\pm$ 0.04 & 0.77 \\
    ExpRubick &\textbf{0.00 $\pm$ 0.02} & 1.10 $\pm$ 0.18 & 0.78 $\pm$ 0.08 & 0.80 $\pm$ 0.12 & \textbf{0.91} \\

 \hline
 
    \end{tabular}
    \caption{Performance Comparison of \textit{DLST-agent} with existing strategies averaged over all the 8 domains (Airport Site, Camera, Energy, Fitness, Flight, Grocery, Itex-Cypress, Outfit - All are readily available in GENIUS). Best Results are in \textbf{bold}.}
    \label{table:hypB}
    \vspace{-4mm}
\end{table*}


\section{Conclusions and future Work}\label{conclusion}
This work uses an actor-critic architecture based deep reinforcement learning to support negotiation in domains with multiple issues. 
In particular, it exploits ``interpretable'' strategy templates used in the state-of-the-art to learn the best combination of acceptance and bidding tactics at any negotiation time, and among its tactics, it uses an adaptive threshold utility, all learned using the DDPG algorithm which derives an initial neural network strategy via supervised learning.
We have empirically evaluated the performance of our DLST-based approach against the ``teacher strategies'' as well as the agent strategies of ANAC'17 and ANAC'18 competitions (since the tournament allowed learning from previous negotiations) in different settings,  showing that our agent outperforms opponents known at training time and can effectively transfer its knowledge to environments with previously unseen opponent agents and domains.

An open problem worth pursuing in the future is how to learn transferable strategies for \textit{concurrent} bilateral negotiations over multiple issues.

\bibliographystyle{ACM-Reference-Format} 
\bibliography{ref}


\end{document}